\documentclass[aps,pra,twocolumn,showpacs,superscriptaddress]{revtex4}
\usepackage{dcolumn}                 
\usepackage{graphicx}
\newcolumntype{w}[1]{D{.}{.}{#1}}

\usepackage{times}
\usepackage{bm}
\usepackage{nicefrac}
\usepackage{amsmath}
\usepackage{amsfonts}
\usepackage{amssymb}
\usepackage{amsthm}

\newcommand{\lbr}{\langle}
\newcommand{\rbr}{\rangle}
\newcolumntype{.}{D{x}{}{-1}}

\newcommand{\SixJ}[6]{
        \left\{
        \begin{array}{ccc}
        #1  & #2  & #3 \\
        #4  & #5  & #6 \\
        \end{array}
        \right\}
        }

\begin{document}
\preprint{Version 0.2}

\title{ Quantum electrodynamic calculation of the hyperfine structure of $^{\bm 3}$He }

\author{K. Pachucki}
\affiliation{Faculty of Physics, University of Warsaw, Ho\.za 69, 
             00-681 Warsaw, Poland}

\author{V. A. Yerokhin}
\affiliation{St. Petersburg State Polytechnical University,
        Polytekhnicheskaya 29, St. Petersburg 195251, Russia}

\author{P. Cancio Pastor}
\affiliation{Istituto Nazionale di Ottica-CNR (INO-CNR), Via Nello Carrara 1,
  I-50019 Sesto Fiorentino, Italy}
\affiliation{ European Laboratory for Non-Linear Spectroscopy (LENS) and
Dipartimento di Fisica,
Universit\'a di Firenze, Via N. Carrara 1, I-50019 Sesto Fiorentino, Italy}

\begin{abstract}
The combined fine and hyperfine structure of the $2^3P$ states in $^3$He is
calculated within the framework of nonrelativistic quantum
electrodynamics. The calculation accounts for the effects of order $m\alpha^6$
and increases the accuracy of theoretical predictions by an order of
magnitude. The results obtained are in good agreement with recent 
spectroscopic measurements in $^3$He. 
\end{abstract}

\maketitle

\section{Introduction}

Spectroscopic measurements of the helium atom have presently reached the level of
accuracy at which they are sensitive to uncertainties of fundamental constants and 
the nuclear effects. This fact can be utilized for improving our
knowledge on fundamental constants and nuclear parameters. 
A recent example is the independent determination
of the fine structure constant $\alpha$ \cite{pachucki:10:hefs} made by
comparison of the theoretical prediction and the experimental 
results for the $^4$He fine structure. Good
agreement of the obtained value of $\alpha$
with the more precise determinations
\cite{hanneke:08,bouchendira:11}  
provided a highly sensitive test of consistency of different theories 
across a wide range of energy scales. 

Recent optical measurements in $^3$He \cite{rooij:11,cancio:12:3he} 
and $^4$He \cite{cancio:04} achieved
the relative precision of about $10^{-11}$ and thus became sensitive to the
uncertainties of the Rydberg constant and the nuclear charge radius.
These experiments created a possibility for the spectroscopic determination of 
the nuclear charge radii of $^3$He and $^4$He, with a significantly higher accuracy 
than can be reached by the electron scattering methods \cite{sick:08:1}. Such
determination is of particular interest now, in 
view of the discrepancy for the proton charge radius 
observed in the muonic hydrogen experiment \cite{pohl:10} and the proposed
follow-up experiment on the muonic helium \cite{antognini:11}. 
Realization of this project requires progress in theory, 
namely a complete calculation of
the $m\,\alpha^7$ corrections to the energy levels. 
Such calculation is difficult but probably feasible,
at least for the low-lying triplet states. 

While the present theory is not accurate enough to provide the nuclear charge
radii of $^3$He and $^4$He separately, it can provide their difference
$\delta r^2$, as the isotope shift is considerably simpler to calculate
than the energy levels. Two such determinations have recently been reported 
by experiments of the Amsterdam
\cite{rooij:11} and the Florence \cite{cancio:12:3he} groups, their results being in
disagreement of four standard deviations. There was also the older spectroscopic
value of $\delta r^2$ by Shiner {\em et al.}~\cite{shiner:95}, which 
relied on the theory of the hyperfine splitting available at that time.

The main
goal of the present investigation is to calculate the combined fine and hyperfine structure
of the $2^3P$ levels of $^3$He with the complete treatment of the $m\alpha^6$
corrections. 
The results obtained 
represent an order-of-magnitude improvement
over the previous theory \cite{morton:06, wu:07} and are
in good agreement with the experimental data available
\cite{cancio:12:3he,smiciklas:03}. The improved theory 
allows us to make a reevaluation of the
$\delta r^2$ determination by Shiner {\em et al.}~\cite{shiner:95}, in reasonable
agreement with the Florence result \cite{cancio:12:3he}.
We also update the previous results 
for the hyperfine splitting of the $2^3S$ state of $^3$He
\cite{pachucki:01:jpb}, making its treatment 
fully consistent with that of the $2^3P$ states.

\section{General approach}

The $2^3P$ energy level in $^3{\rm He}$ is split by the hyperfine and fine
structure effects. The hyperfine splitting is induced by the interaction 
between the dipole magnetic moment of the nucleus
and that of the electrons, whereas the fine structure is due to the interaction
between the electron spin and the electron orbital angular momentum. 
For heavy atoms, it is common that the hyperfine splitting 
(being suppressed by the electron-to-proton mass ratio) 
is much smaller than the fine structure splitting, and so 
the hyperfine and fine structure effects can be investigated
separately. For the $2^3P$ states of $^3{\rm He}$, however, both effects are of
the same order of magnitude and thus should be calculated together. 

The general method of calculation of the combined fine and hyperfine
structure
is the nonrelativistic perturbation theory for quasidegenerate
states. The active space of (strongly interacting) quasidegenerate states is defined
as $2^3P_J^F$, where $F$ is the total angular momentum of the atomic state and
$J = 0$, 1, 2 is the electronic angular momentum. In the case of $^3{\rm He}$,
the nuclear spin $I=\nicefrac12$, so $F=J\pm \nicefrac12>0$ and the active space
consists of five levels. Note that, unlike in previous studies
\cite{marin:95,morton:06, wu:07}, we do not include the 
$2^1P$ state in the active space, as its mixing with the $2^3P$
levels is relatively weak. The contribution of this state is accounted for by 
the standard perturbation theory for non-degenerate states.

The energies of the $2^3P_J^F$ states are the eigenvalues of the $5\times 5$ matrix 
of the effective Hamiltonian $H$, whose
elements are
\begin{align} \label{e1}
E^F_{J,J'} \equiv \lbr FJM_F| H |FJ'M_F\rbr\,,
\end{align}
where $M_F$ is the projection of the total angular momentum $F$. (Since the energies do not
depend on $M_F$, it can be fixed arbitrary.)
In practical calculations, it is convenient to consider the shifts of
the individual $2^3P_J^F$ levels with respect to the $2^3P$ centroid. In other words, we
require that
\begin{align}\label{ce1}
\sum_F (2\,F+1) \sum_J E^F_{J,J}=0\,.
\end{align}
In this case, all effects that do not depend on the nuclear and/or electron spin
do not contribute and can be omitted in the calculations.

The matrix elements of the effective Hamiltonian (\ref{e1}) are obtained
in this work by expansion in terms of the fine structure constant $\alpha$,
\begin{align} 
E^{F}_{J,J'} = &\ 
{\langle H_{\rm fs} \rangle}_J\, \delta_{J,J'}+
\langle H^{(4+)}_{\rm hfs}\rangle + 
\langle H^{(6)}_{\rm hfs} \rangle 
  \nonumber \\ &
+ 
2\,\langle H_{\rm hfs}^{(4)}\,\frac{1}{(E-H)'}\,
\bigl[H_{\rm nfs}^{(4)}+H_{\rm fs}^{(4)}\bigr]\rangle
  \nonumber \\ &
+ 
\langle H_{\rm hfs}^{(4)}\,\frac{1}{(E-H)'}\,H_{\rm hfs}^{(4)}\rangle
+ \langle H_{\rm nucl\& ho} \rangle\,, \label{01}
\end{align}
where $H_{\rm fs}$ is the effective operator responsible for the
fine-structure splitting in the absence of the nuclear spin and the other
terms are the nuclear-spin-dependent contributions. 
$H^{(4+)}_{\rm hfs}$ 
is the leading hyperfine Hamiltonian with the
recoil and anomalous magnetic moment additions; it is of nominal order
$m\alpha^4$ and contains parts of higher-order contributions. $H^{(6)}_{\rm hfs}$
is the effective Hamiltonian of order $m\alpha^6$ and is derived in the
present work. The next two terms in Eq.~(\ref{01}) are the second-order
corrections that also contribute to order $m\alpha^6$. $H^{(4)}_{\rm hfs}$ 
is $H^{(4+)}_{\rm hfs}$ without the
recoil and anomalous magnetic moment additions, $H^{(4)}_{\rm fs}$ is
the effective Hamiltonian responsible for the leading fine-structure splitting
of order $m\alpha^4$,
and  $H^{(4)}_{\rm nfs}$ is
the effective Hamiltonian responsible for spin independent effects 
of order $m\alpha^4$. Finally, $H_{\rm nucl\& ho}$ represents the nuclear
effects and the higher-order $(\sim m\alpha^7)$ QED corrections proportional to the
$\delta$ function at the origin. This part cannot be accurately calculated at
present, because of insufficient theoretical knowledge of the nuclear structure.
It will be obtained from the experiment on the ground-state
hyperfine splitting in $^3{\rm He}$. 

In order to facilitate the evaluation of the matrix elements in
Eq.~(\ref{01}), it is convenient to factorize out their dependence on the
nuclear degrees of freedom $F$ and $M_F$ and on the electronic angular momentum
$J$. It can be achieved by observing that any operator contributing to
Eq.~(\ref{01}) can be represented in terms of six basic angular-momentum
operators: $(\vec S\cdot\vec L)$, $(\vec I\cdot\vec L)$, $(\vec I\cdot\vec S)$,
$(S^i\,S^j)^{(2)}\,(L^i\,L^j)^{(2)}$, $I^i\,S^j\,(L^i\,L^j)^{(2)}$, and
$I^i\,L^j\,(S^i\,S^j)^{(2)}$, where the second-order tensors are defined by
$(L^i\,L^j)^{(2)} \equiv  \nicefrac12\, L^iL^j+\nicefrac12\, L^jL^i
-\nicefrac13\, \vec L^2\delta^{ij}$ and the summation over the repeated
indices is assumed. Since the nuclear spin of helion $I = \nicefrac12$, there
are no operators quadratic in $I$. So, Eq.~(\ref{01}) 
can be represented as
\begin{eqnarray}
E^F_{J,J'} &=& A_{\rm sl}\,\lbr\vec S\cdot\vec L\rbr +
A_{\rm ss}\,\lbr(S^i\,S^j)^{(2)}\,(L^i\,L^j)^{(2)}\rbr
\nonumber \\ &&
+ A_{\rm s} \lbr\vec I\cdot\vec S\rbr
+ A_{\rm l} \lbr\vec I\cdot\vec L\rbr
+ A_{\rm s l l}\,\lbr I^i\,S^j\,(L^i\,L^j)^{(2)}\rbr
\nonumber \\ &&
+ A_{\rm s s l}\,\lbr I^i\,L^j\,(S^i\,S^j)^{(2)}\rbr\,, \label{03}
\end{eqnarray}
where the constants $A_i$ do not depend on $F$, $M_F$, $J$, $J'$, and are
represented by expectation values of purely electronic operators between the
spatial $2^3P$ wave functions. The
matrix elements of the basic angular-momentum operators are calculated
analytically by means of the Racah algebra and listed in Appendix~\ref{app:A}.

The first two terms in the right-hand-side of Eq.~(\ref{03}) do not depend on
the nuclear spin and correspond to the fine structure.
The fine structure splitting in the absence of the nuclear spin was investigated in
our previous investigations \cite{pachucki:09:hefs,pachucki:10:hefs},
so we use results obtained there. 

\section{Leading hyperfine structure}
The leading hyperfine-structure Hamiltonian $H^{(4+)}_{\rm hfs}$ is well
known. For our purposes, it is convenient to write it in the following form,
\begin{align}
H^{(4+)}_{\rm hfs} = &\ \frac{m_r^3}{m\,M}(1+\kappa)\,\alpha^4\,
\biggl[\vec I\cdot \vec S\,Q + \vec I\cdot \vec Q + I^i\, S^j\,Q^{ij}
 \nonumber \\ &
+ \vec I\cdot\vec S_A\,Q_A + I^i\,S_A^j\,Q_A^{ij}\biggr]\,, \label{09}
\end{align}
where 
$\vec S$ and $\vec S_A$ are the electron spin operators,
$\vec S = (\vec\sigma_1+\vec\sigma_2)/2$ and $\vec S_A = (\vec \sigma_1-\vec \sigma_2)/2$,
$m_r$ is the reduced mass of the electron-nucleus system, $M$ is the
nuclear mass, and 
$\kappa$ is the magnetic moment anomaly related to the nuclear dipole
magnetic moment $\mu$ by
\begin{align}
\frac{m}{M}(1+\kappa) \equiv \frac{\mu}{\mu_B}\,\frac{1}{2 Z I}\,,
\end{align} 
where $\mu_B$ is the Bohr magneton.
The electronic operators are given by (in atomic units)
\begin{align}
Q &\ =   (1+a_e)\, \frac{Z}{3}\, 4\pi \bigl[\delta^3(r_1)+\delta^3(r_2)\bigr]\,,
  \\
Q_A &\ = (1+a_e)\, \frac{Z}{3}\, 4\pi \bigl[\delta^3(r_1)-\delta^3(r_2)\bigr]\,,
\end{align}
\begin{align}
\vec{Q} = &\  Z\, \left[
  \frac{\vec{r}_1}{r_1^3}\times \vec{p}_1 + \frac{\vec{r}_2}{r_2^3}\times
  \vec{p}_2\right]
   \nonumber \\ &
    + \frac{m}{M}\,\frac{1+2\kappa}{1+\kappa}\, \frac{Z}{2}\,
   \left[
  \frac{\vec{r}_1}{r_1^3} + \frac{\vec{r}_2}{r_2^3}\right]\times
  (\vec{p}_1+\vec{p}_2)\,,
\end{align}
\begin{align}
Q^{ij} &\ = -(1+a_e)\, \frac{Z}{2}\,\biggl[
            \frac1{r_1^3}\biggl( \delta^{ij}-3\frac{r_1^ir_1^j}{r_1^2}\biggr)
         + \frac1{r_2^3}\biggl( \delta^{ij}-3\frac{r_2^ir_2^j}{r_2^2}\biggr)
   \biggr]\,,
 \\
Q^{ij}_A &\ = -(1+a_e)\, \frac{Z}{2}\,\biggl[
           \frac1{r_1^3}\biggl( \delta^{ij}-3\frac{r_1^ir_1^j}{r_1^2}\biggr)
         - \frac1{r_2^3}\biggl( \delta^{ij}-3\frac{r_2^ir_2^j}{r_2^2}\biggr)
   \biggr]\,,
\end{align}
where $a_e$ is the anomalous magnetic moment of the electron.

The matrix element of the Hamiltonian 
between the $2^3P_J^F$ states can be represented as
\begin{align}
{\langle H^{(4+)}_{\rm hfs}\rangle}_{JJ'}^F = &\ 
  \frac{m_r^3}{m\,M}(1+\kappa)\,\alpha^4\,\biggl[
 \langle ^3\vec{P}|Q|^3\vec{P}\rangle\,\lbr\vec I\cdot\vec S\rbr
 \nonumber \\ &
+\frac{1}{2}\,\langle ^3\vec{P}|\vec{Q}|^3\vec{P}\rangle\, \lbr\vec I\cdot \vec L\rbr
 \nonumber \\ &
-\frac{3}{5}\,\langle ^3\vec{P}|\hat{Q}|^3\vec{P}\rangle\, \lbr I^i\,S^j\,(L^i\,L^j)^{(2)}\rbr
  \biggr]\,, \label{e2}
\end{align}
where the shorthand notations for the
matrix elements of the electronic operators are 
\begin{eqnarray} 
\langle\vec\phi|Q|\vec\psi\rangle &=& \langle\phi_i|Q|\psi_i\rangle\,, \\
\langle\vec\phi|\vec Q|\vec\psi\rangle &=& -i\,\epsilon^{ijk}\langle\phi_i|Q_j|\psi_k\rangle\,,\\
\langle\vec\phi|\hat Q|\vec\psi\rangle &=& \langle\phi_i|Q_{ij}|\psi_j\rangle\,,
\end{eqnarray}
and the spatial triplet odd $P$ wave function is represented as
\begin{align}
\psi^i(r_1,r_2,r) = r_1^i\,f(r_1,r_2,r) - (r_1 \leftrightarrow r_2)\,,
\end{align}
with $f(r_1,r_2,r)$ being a real scalar function of $r_1$, $r_2$, and $r\equiv
|\vec{r}_1-\vec{r}_2|$.
The $P$-state wave function defined above is real and normalized to
$\langle \vec P|\vec P\rangle = \langle P_i|P_i\rangle=1$.

Matrix elements of the operators $Q$, $\vec{Q}$, and $\hat{Q}$ in 
Eq.~(\ref{e2}) are evaluated between the
wave functions that include the mass polarization term in the nonrelativistic
Hamiltonian. 
Numerical results for the expectation values 
of these operators are presented in Table~\ref{tab:leading}.
\begin{table}
\caption{
Expectation values of the $m\alpha^4$ hyperfine operators for the $2^3P$ state
of $^3{\rm He}$, with the mass polarization included, in a.u.
The expected numerical uncertainty is less than 1 on the last significant
digit. 
\label{tab:leading}
} 
\begin{ruledtabular}
  \begin{tabular}{l.}
$Q$  &  \multicolumn{1}{c}{$\langle Q\rangle$}    \\
     \hline\\[-5pt]
$\frac{\displaystyle 2\,Z}{\displaystyle  3}\, 4\pi \,\delta^3(r_1)$
&  21x.092\,193  \\
$2\,Z\,\frac{\displaystyle \vec{r}_1}{\displaystyle r_1^3}\times \vec{p}_1$
&   0x.277\,443  \\
$ Z\,\left(\frac{\displaystyle \vec{r}_1}{\displaystyle r_1^3} + \frac{\displaystyle \vec{r}_2}{\displaystyle r_2^3}\right) \times \vec{p}_1$
&  -0x.060\,491  \\
$-Z\,\frac{\displaystyle 1}{\displaystyle r_1^3}\biggl( \delta^{ij}-3\frac{\displaystyle r_1^ir_1^j}{\displaystyle r_1^2}\biggr)$
&   0x.140\,325  \\
  \end{tabular}
\end{ruledtabular}
\end{table}
We observe that the Fermi contact interaction yields a dominating
contribution, which could be aniticipated, as it comes from
both $1s$ and $2p$ electrons. This also explains why the hyperfine splitting
is of the same order as the fine splitting, since the latter 
comes mainly from the $2p$ electron.

\section{${\bm m\bm \alpha^{\bm 6}}$ corrections}
\label{sec:ma6}

Calculation of the $m\alpha^6$ contribution to the hyperfine structure of 
the $2^3 P$ state is the principal objective of this work. Analogous calculation
for the $2^3S_1$ state have already been performed in
Ref.~\cite{pachucki:01:jpb}. Here we verify the calculation for the $2^3S$
state and extend it to the $2^3P$ state.

Derivation of the effective Hamiltonian to order $m\alpha^6$ for an arbitrary
state of helium is given in Appendix~\ref{app:B}. The result is
\begin{align} \label{eH6}
H^{(6)}_{\rm hfs} = &\ \frac{m_r^3}{m\,M}\,(1+\kappa)\,\alpha^6\,\Bigl[
\vec I\cdot \vec S\,(P_{\rm rad}+P_{\rm nrad}) 
 \nonumber \\ &
+ \vec I\cdot \vec P + I^i\, S^j\,P^{ij}\Bigr]
\end{align}
where 
\begin{eqnarray}
P_{\rm rad} = 
Z\,\biggl(\ln2-\frac{5}{2}\biggr)\,\frac{Z}{3}\,4\pi\left[\delta^3(r_1)+\delta^3(r_2)\right]\,,
\end{eqnarray}
is the effective operator induced by the radiative QED effects,
\begin{align}
P_{\rm nrad} = &\ \frac{Z^2}{3}\,\frac{1}{r_1^4}
-\frac{Z}{3}\,p_1^2\,4\,\pi\,\delta^3(r_1)
 \nonumber \\ &
- Z\,\frac{\vec r}{r^3}\cdot\frac{\vec r_1}{r_1^3}
+\frac{8}{3}\,\biggl(\ln 2 - \frac{5}{2}\biggr)\,Z^2\,\pi\,\delta^3(r_1)\,,
\end{align}
\begin{align}
\vec P = -
Z\,p_1^2\,\frac{\vec r_1}{r_1^3}\times\vec p_1
-Z\,\frac{\vec r_1}{r\,r_1^3}\,\times\vec p_2 
-Z\,\biggl(\frac{\vec r_1}{r_1^3}\times\frac{\vec r}{r^3}\biggr)\,
(\vec r\cdot\vec p_2)\,, 
\label{60}
\end{align}
and
\begin{align}
P^{ij}=&\ -\frac{Z}{2}\biggl(\frac{Z}{3\,r_1}+p_1^2\biggr)\,
\biggl(3\,\frac{r_1^i\,r_1^j}{r_1^{\;5}}-\frac{\delta^{ij}}{r_1^{\;3}}\biggr)
  \nonumber \\ &
+\frac{Z}{2}\,\biggl(3\,\frac{r^j}{r^3}\,\frac{r_1^i}{r_1^3}
-\delta^{ij}\,\frac{\vec r}{r^3}\cdot\frac{\vec
  r_1}{r_1^3}\biggr)\,.\label{61}
\end{align}
It can be immediately seen that the operator $P_{\rm nrad}$ involves 
highly singular operators, whose matrix elements between the $^3P$ functions 
diverge. However, the divergence cancels 
out if $P_{\rm nrad}$ is considered together with the second-order $m\alpha^6$ 
contribution, as is explained below.

The second-order $m\alpha^6$ correction $\delta E_{\rm sec}$ is given by
\begin{align} \label{e123}
\delta E_{\rm sec} = 2\,\langle H_{\rm hfs}^{(4)}\,\frac{1}{(E-H)'}\,
\bigl[H_{\rm nfs}^{(4)}+H_{\rm fs}^{(4)}\bigr]\rangle \,,
\end{align}
where $H_{\rm hfs}^{(4)}$ is obtained from $H_{\rm hfs}^{(4+)}$ from
Eq. (\ref{09}) by dropping the  electron magnetic moment anomaly $a_e$ and
the recoil part, and 
\begin{align}
H^{(4)}_{\rm nfs} + H^{(4)}_{\rm fs} = T + \vec S\cdot\vec T + S^i\,S^j\,T^{ij}
+ \vec S_A\cdot\vec T_A\,,
\end{align}
where $H^{(4)}_{\rm nfs} \equiv T$ is the spin independent part of 
the effective Hamiltonian of order $m\alpha^4$,
\begin{align}
T = &\  -\frac{p_1^4+p_2^4}{8}
  + \frac{Z\pi}{2}\,\left[ \delta^3(r_1)+\delta^3(r_2)\right]
 \nonumber \\ &
  -\frac12\,p_1^i\biggl( \frac{\delta^{ij}}{r}+\frac{r^ir^j}{r^3}\biggr)p_2^j\,,
\end{align}
and the remaining operators are responsible for the fine structure to order $m\alpha^4$,
\begin{align}
 \vec{T} = &\ \frac{Z}{4}\, \left[ \frac{\vec{r}_1}{r_1^3}\times \vec{p}_1
  + \frac{\vec{r}_2}{r_2^3}\times \vec{p}_2 \right]
  + \frac{3}{4}\,\frac{\vec r}{r^3}\times (\vec{p}_2-\vec{p}_1)\,,
  \\
 \vec{T}_A = &\ \frac{Z}{4}\, \left[ \frac{\vec{r}_1}{r_1^3}\times \vec{p}_1
  - \frac{\vec{r}_2}{r_2^3}\times \vec{p}_2 \right]
  + \frac{1}{4}\, \frac{\vec{r}}{r^3} \times (\vec{p}_2+\vec{p}_1)\,,
\end{align}
and
\begin{align}
 T^{ij} = &\ \frac12\, \frac1{r^3}\left( \delta^{ij} - 3\frac{r^ir^j}{r^2}\right)\,.
\end{align}

\begin{table}
\caption{
Expectation values of the $m\alpha^6$ hyperfine operators for the $2^3P$ state
of $^3{\rm He}$, in a.u.
The expected numerical uncertainty is less than 1 on the last significant
digit. 
\label{tab:ma6first}
} 
\begin{ruledtabular}
  \begin{tabular}{l.}
$P$  &  \multicolumn{1}{c}{$\langle P\rangle$}    \\
     \hline\\[-5pt]
$P_{\rm rad}$  & -76.x220\,978 \\
$P^{\prime}_{\rm nrad}$ &   0.x367\,674 \\ 
$\vec{P}          $   &  -0.x529\,385 \\
$\hat{P}$             &  -0.x361\,070 
  \end{tabular}
\end{ruledtabular}
\end{table}

\begin{table}
\caption{
Individual second-order matrix elements, for different types of intermediate
states $^sL$, in a.u.
The expected numerical uncertainty is less than 1 on the last significant
digit, when not given explicitly.
\label{tab:ma6sec}
} 
\begin{ruledtabular}
  \begin{tabular}{ll.}
$^sL$
&  $(A,B)$ & \multicolumn{1}{c}
 {$\left\langle A\, \frac{\displaystyle 1}{\displaystyle
     (E-H)'}B\,\right\rangle$} \\
     \hline\\[-5pt]
$^3P$ & $({Q'}$ , ${T'})$  &      63.x531\,80 \\ 
      & $(\vec{Q}$  , ${T})$  &    0.x085\,734   \\ 
      & $(\hat{Q}$  , ${T})$  &    0.x044\,830 \\ 
      & $({Q}$  , $\vec{T})$  &    0.x030\,473  \\
      & $(\vec{Q}$  , $\vec{T})$&  0.x010\,782  \\    
      & $(\hat{Q}$  , $\vec{T})$& -0.x051\,433 \\ 
      & $({Q}$  , $\hat{T})$  &   -0.x048\,826  \\ 
      & $(\vec{Q}$  , $\hat{T})$&  0.x237\,962 \\ 
      & $(\hat{Q}$  , $\hat{T})$&  0.x114\,340 \\ 
\hline
$^1P$ & $({Q_A}$ , $\vec{T}_A)$  & 157.x531\,64 \\ 
      &$(\hat{Q}_A$ , $\vec{T}_A)$& -1.x150\,(5) \\
\hline
$^3D$ & $(\vec{Q}$ , $\vec{T})$  &  0.x001\,148 \\
      & $(\vec{Q}$ , $\hat{T})$  & -0.x000\,434 \\  
      & $(\hat{Q}$ , $\vec{T})$  &  0.x020\,6\,(2) \\
      & $(\hat{Q}$ , $\hat{T})$  & -0.x000\,740 \\ 
\hline
$^1D$ & $(\hat{Q}_A$ , $\vec{T}_A)$& 0.x001\,8\,(3) \\
\hline
$^3F$ & $(\hat{Q}$ , $\hat{T})$  &  0.x011\,100  \\ 
  \end{tabular}
\end{ruledtabular}
\end{table}

Let us now consider the sum of singular contributions,
\begin{align}
\delta_{\rm sing} E = &\ \frac{m_r^3}{m\,M}\,(1+\kappa)\,\alpha^6\,\vec I\cdot\vec S\,\biggl[
\langle P_{\rm nrad}\rangle 
  \nonumber \\ &
+ \biggl\langle Q\,\frac{1}{(E-H)'}\,T + {\rm h.c.}\biggr\rangle\biggr]\,.
\end{align}
While the expectation values of the operators $Q$ and $T$ are
finite, the second-order matrix element of these operators diverges.
In order to eliminate the divergences,
we regularize the Coulomb electron-nucleus interaction by introducing 
an artificial  parameter $\lambda$,
\begin{align}
\frac{Z}{r_a}\rightarrow \frac{Z}{r_a}\,
\bigl(1-e^{-\lambda\,m\,Z\,r_a}\bigr)\,,
\end{align}
All other electron-nucleus interaction terms are  
regularized in the same way.
This entails the following replacements in effective Hamiltonians,
\begin{align}
4\,\pi\,Z\,\delta^3(r_a) \equiv&\ -\nabla^2\,\frac{Z}{r_a}
\rightarrow -\nabla^2\,\frac{Z}{r_a}
\bigl(1-e^{-\lambda\,m\,Z\,r_a}\bigr)\,, \\
\frac{Z^2}{r_a^4} \equiv&\ 
\biggl(\vec\nabla\frac{Z}{r_a}\biggr)^2 \rightarrow
\biggl[\vec\nabla\frac{Z}{r_a}
\bigl(1-e^{-\lambda\,m\,Z\,r_a}\bigr)\biggr]^2\,.
\end{align}
Once the 
electron-nucleus interaction is regularized, one can in principle calculate
all matrix elements and take the limit $\lambda\rightarrow\infty$.
However, since matrix elements can be calculated only numerically,
it is more convenient to transform the effective operators
to the regular form, where $\lambda$ can be taken to infinity
before numerical calculations. 

To this end, we transform the operators in the second-order matrix element
$T\to T'$ and $Q\to Q'$ by
\begin{eqnarray} \label{e125}
T' &\equiv& T
-\frac{1}{4}\,\sum_a\,\biggl\{\frac{Z}{r_a}\,,\,E-H\biggr\}\,, \\
Q' &\equiv& Q+\frac{2}{3}\,\sum_a\,\biggl\{\frac{Z}{r_a}\,,\,E-H\biggr\}\,,
 \label{e126}
\end{eqnarray}
where $\left\{\ldots,\ldots\right\}$ is the commutator and the implicit 
$\lambda$-regularization is assumed.
After this transformation, the singular part takes the form
\begin{align}
\delta_{\rm sing} E = &\ \frac{m_r^3}{m\,M}\,(1+\kappa)\,\alpha^6\,\vec I\cdot\vec S\,\biggl[
\langle P'_{\rm nrad}\rangle 
 \nonumber \\ & 
+ \biggl\langle Q'\,\frac{1}{(E-H)'}\,T' + {\rm h.c.}\biggr\rangle\biggr]
\end{align}
where both the first and second order terms are separately finite
and thus the limit $\lambda\rightarrow\infty$ can be evaluated analytically. The result
for the regularized operator $P'_{\rm nrad}$ is
\begin{widetext}
\begin{eqnarray}
\langle P'_{\rm nrad}\rangle  &=& \frac23\,\biggl\langle \biggl(E-\frac{1}{r}\biggr)^2\,
\biggl(\frac{Z}{r_1}+\frac{Z}{r_2}\biggr)+
\biggl(E-\frac{1}{r}\biggr)\,
\biggl(\frac{Z^2}{r_1^2}+\frac{Z^2}{r_2^2}+
4\,\frac{Z}{r_1}\,\frac{Z}{r_2}\biggr)
\nonumber \\ &&
+2\,\frac{Z}{r_1}\,\frac{Z}{r_2}\,\biggl(\frac{Z}{r_1}+\frac{Z}{r_2}\biggr)
-\biggl(E-\frac{1}{r}+\frac{Z}{r_2}-
\frac{p_2^2}{2}\biggr)\,4\,\pi\,Z\,\delta^3(r_1)-
\frac{5\,Z}{4}\,\frac{r^i}{r^3}\,\biggl(
\frac{r^i_1}{r_1^3}-\frac{r^i_2}{r_2^3}\biggr)
\nonumber \\ &&
+p^i_1\,\frac{Z^2}{r_1^2}\,p^i_1 
+p_2^2\,\frac{Z^2}{r_1^2}
-p_2^2\,\frac{Z}{r_1}\,p_1^2
+2\,p_2^i\,\frac{Z}{r_1}\,\biggl(
\frac{\delta^{ij}}{r}+\frac{r^i\,r^j}{r^3}\biggr)\,p_1^j\biggr\rangle
\nonumber \\ &&
-\frac{1}{6}\,\biggl\langle\frac{Z}{r_1}+\frac{Z}{r_2}\biggr\rangle\,
\langle 4\,\pi\,Z\,[\delta^3(r_1)+\delta^3(r_2)]\rangle
+\frac43\,\biggl\langle\frac{Z}{r_1}+\frac{Z}{r_2}\biggr\rangle\,
\langle T \rangle
\,.
\end{eqnarray}
\end{widetext}
Matrix elements of the regularized $m\alpha^6$ Hamiltonian
are obtained by
\begin{align}
\langle H^{(6)}_{\rm hfs} \rangle &\ = 
\frac{m_r^3}{m\,M}\,(1+\kappa)\,\alpha^6\,\biggl[
\langle ^3\vec P|\left(P_{\rm rad}+P^{\prime}_{\rm nrad}\right)|^3\vec P\rangle\,\lbr \vec I\cdot\vec S \rbr
 \nonumber \\ &
+\frac{1}{2}\,\langle ^3\vec P|\vec P|^3\vec P \rangle\,\lbr \vec I\cdot \vec L\rbr
 \nonumber \\ &
-\frac{3}{5}\,\langle ^3\vec P|\hat P|^3\vec P \rangle\,\lbr I^i\,
S^j\,(L^i\,L^j)^{(2)}\rbr
\biggr]\,.
\end{align}
The numerical results for the expectation values of the electronic operators
are listed in Table~\ref{tab:ma6first}. 

The second-order $m\alpha^6$ correction [Eq.~(\ref{e123})] is conceptually
simple but its calculation is
rather involved technically. This is partly because of 
the coupling with
intermediate states of different symmetries ($^3P$, $^1P$, $^3D$, $^1D$, and $^3F$) and
partly due to the presence of nearly singular operators. For achieving
numerically stable results, some operators had to be transformed to the
regular form by using the same method as for the singular
part. Details of the calculation and the regularization procedure are
described in Appendix \ref{app:C}. 

Numerical results for the second-order $m\alpha^6$ corrections are presented in
Table~\ref{tab:ma6sec}. Note a large contribution of the $^1P$ intermediate
states, which is mainly due to the $2^3P$-$2^1P$ mixing. We also mention that
the numerical uncertainties of the second-order corrections are completely
negligible on the level of the total theoretical error.

\section{Second-order HFS correction}
\label{sec:2ndhfs}

The fifth term in Eq.~(\ref{01}) is the second-order hyperfine correction. 
Its nominal order is $(m^2/M)\,\alpha^6$. However, it is enhanced by the
small $2^3P$-$2^1P$ energy difference in the denominator, which 
makes it numerically significant. 
Rigorous calculation of this contribution is difficult because of
divergences, which are canceled by the corresponding contribution from the
nuclear-structure effects. 
In the present work, we approximate the
second-order hyperfine contribuiton by keeping only the lowest
lying $^1P^F_1$ intermediate state in the sum over spectrum. 

The second-order hyperfine correction induces 
contributions to the isotope shift (i.e., to the centroid of the $2^3P$ level),
\begin{widetext}
\begin{align} \label{eq002}
\langle H_{\rm hfs}^{(4)}\,\frac{1}{(E-H)'}\,{H_{\rm hfs}^{(4)}\rangle}_{\rm iso}
\approx \frac{m_r^5}{m^2\,M^2}\,(1+\kappa)^2\,\alpha^6\,
\frac{1}{E(2^3P)-E(2^1P)}\,\biggl[
\frac{1}{4}\, \langle2^3\vec P|Q_A|2^1\vec P\rangle^2
+\frac{1}{20}\,
\langle2^3\vec P|\hat Q_A|2^1\vec P\rangle^2\,
\biggr]\,,
\end{align}
to the fine structure,
\begin{align} 
\langle H_{\rm hfs}^{(4)}\,\frac{1}{(E-H)'}\,{H_{\rm hfs}^{(4)}\rangle}_{\rm fs}
\approx & \frac{m_r^5}{m^2\,M^2}\,(1+\kappa)^2\,\alpha^6\,\frac{1}{E(2^3P)-E(2^1P)}\,\biggl[
\langle2^3\vec P|Q_A|2^1\vec P\rangle\,\langle2^3\vec P|\hat Q_A|2^1\vec P\rangle\,
\lbr \frac{3}{10}\,(S^i\,S^j)^{(2)}\,(L^i\,L^j)^{(2)}\rbr
\nonumber \\&
+\langle2^3\vec P|\hat Q_A|2^1\vec P\rangle^2\,\lbr
 -\frac{9}{160}\,\vec L\cdot\vec S
+\frac{21}{400}\,(L^i\,L^j)^{(2)}\,(S^i\,S^j)^{(2)}
\rbr \biggr]\,, \label{59}
\end{align}
and to the hyperfine structure,
\begin{align} \label{eq004}
\langle H_{\rm hfs}^{(4)}\,\frac{1}{(E-H)'}\,{H_{\rm hfs}^{(4)}\rangle}_{\rm hfs}
\approx & \frac{m_r^5}{m^2\,M^2}\,(1+\kappa)^2\,\alpha^6\,\frac{1}{E(2^3P)-E(2^1P)}\,\biggl[
\langle2^3\vec P|Q_A|2^1\vec P\rangle^2\,\lbr-\frac12\, \vec I\cdot\vec S\rbr
\nonumber \\&
+\langle2^3\vec P|Q_A|2^1\vec P\rangle\,\langle2^3\vec P|\hat Q_A|2^1\vec P\rangle\,
\lbr-\frac{3}{10}\,I^i\,S^j\,(L^i\,L^j)^{(2)}\rbr
\nonumber \\&
+\langle2^3\vec P|\hat Q_A|2^1\vec P\rangle^2\,\lbr
-\frac{3}{40}\,\vec I\cdot\vec L
+\frac{1}{20}\,\vec I\cdot\vec S
+\frac{21}{200}\,I^i\,S^j\,(L^i\,L^j)^{(2)}
-\frac{9}{200}\,I^i\,L^j\,(S^i\,S^j)^{(2)}\rbr\biggr]\,.
\end{align}
\end{widetext}
Numerical results for the second-order hyperfine corrections are presented in
Table~\ref{tab:hfssec}.

The approximate treatment of the second-order hyperfine
contribution yields, in our opinion, the dominant uncertainty of our
theoretical predictions for the hyperfine structure. 
It is, therefore, important to
estimate the neglected part. To this end, we introduce a different
approximation for the second-order hyperfine contribution, which is obtained from 
Eq.~(\ref{eq004}) by the following substitution
\begin{align} \label{eq005}
\frac1{E(2^3P)-E(2^1P)} & \langle2^3\vec P|Q_A|2^1\vec P\rangle^2 
  \to 
\nonumber \\ &
\sum_n \frac1{E(2^3P)-E(n^1P)} \langle2^3\vec P|Q_A'|n^1\vec P\rangle^2\,,
\end{align}
where $Q_A'$ is the regularized $\delta$ function operator given by Eq.~(\ref{Qap}).
The difference between these two approximations (of about a half percent) is
used as the estimated error of the theoretical hyperfine structure.

\begin{table}
\caption{
Second-order hyperfine matrix elements, in a.u.,
all figures shown are exact.
Notations are: $(A,B) = \left\langle 2^3P| A | 2^1P\right\rangle \, 
\left\langle 2^1P| B| 2^3P\right\rangle/[E(2^3P)-E(2^1P)].$
The expected numerical uncertainty is less than 1 on the last significant
digit.
\label{tab:hfssec}
} 
\begin{ruledtabular}
  \begin{tabular}{l.}
 $({Q_A}$ , $Q_A)$             & -47\,774.x980 \\
 $({Q_A}$ , $\hat{Q}_A)$       &      249.x528 \\
 $({\hat{Q}_A}$ , $\hat{Q}_A)$ &       -1.x303
  \end{tabular}
\end{ruledtabular}
\end{table}


\section{Nuclear-structure contribution}

The last term in Eq.~(\ref{01}) is the nuclear-structure and higher-order
QED contribution. 
Accurate calculation of the nuclear contribution is presently not possible 
due to insufficient knowledge of the nuclear dynamics.
However, one may claim that the dominant part of this
effect comes from a local operator proportional to the $\delta$ function at
the origin. 
Using this assumption, one can infer the nuclear-structure correction
in neutral $^3$He
from the experimental value of the ground-state hyperfine splitting
in $^3$He$^+$, measured very
accurately by Schl\"usser {\em et al} \cite{schluesser:69} half a century
ago. 
This approach has been used also in the previous studies of the $^3$He 
hyperfine structure \cite{morton:06,wu:07}. 

In order to infer the nuclear-structure contribution, 
we subtract from the $^3$He$^+$ experimental result the hydrogenic
limit of all corrections accounted for in the previous sections.
The remainder is the nuclear-structure contribution plus some 
(small) higher-order QED corrections. 
Specifically, we define the nuclear-structure and higher-order QED contribution $C_{\rm
  nucl\& ho}$ as
\begin{align}
E_{\rm exp}(1s,^3{\rm He}^+) &\ = \frac{m_r^3}{m\,M}(1+\kappa)\,\alpha^4\,
  \biggl\{ 1+a_e 
 \nonumber \\ &
+ \alpha^2\biggl[\frac32\,Z^2 + Z\biggl( \ln 2-\frac52\biggr) \biggr]
 \nonumber \\ & 
+ \alpha\,C_{\rm nucl\& ho}\biggr\}\,
\frac{2Z}{3}\,
  \langle 4\pi\delta^3(r)\rangle \,,
\end{align} 
where $E_{\rm exp}(1s,^3{\rm He}^+)/h = 8665649.867\,(10)$~kHz is the
experimental result of Ref.~\cite{schluesser:69}. The above equation yields
\begin{align}
C_{\rm nucl\& ho} = -0.031\,891\,.
\end{align} 
This value does not have any theoretical uncertainty
(by definition) and thus is accurate to all figures given. 
The corresponding contribution to the hyperfine structure in neutral helium
then is 
\begin{align}
\langle H_{\rm nucl\& ho} \rangle &\ = 
  \frac{m_r^3}{m\,M}(1+\kappa)\,\alpha^5\, C_{\rm nucl\& ho}\,
 \nonumber \\ & \times
\langle 2^3\vec{P}|\frac{Z}{3}\, 4\pi \bigl[
   \delta^3(r_1)+\delta^3(r_2)\bigr]|2^3\vec{P}\rangle\,
    \langle \vec I\cdot\vec S \rangle 
\,.
\end{align}

\section{Small corrections}

Here we pick up some higher-order contributions, which 
are not accounted for by $C_{\rm nucl\& ho}$
but might be relevant for comparison with
experiment. Since the mixing of the $2^3P$ and $2^1P$ levels by the
fine-structure operator is large, we calculate here the anomalous magnetic
moment and recoil corrections to this mixing, which are of orders 
$\alpha^7\,m^2/M$ and $\alpha^6\,m^3/M^2$, respectively.

The nuclear spin dependent $2^3P$-$2^1P$ fine-structure mixing is given by
\begin{widetext}
\begin{eqnarray} \label{e124}
\langle H_{\rm mix} \rangle &=& 
\frac{m_r^3}{m\,M}\,(1+\kappa)\,\alpha^6\,\frac{m_r^2}{m^2}\,
\frac{1}{E(2^3P)-E(2^1P)}\biggl[
\langle 2^3\vec P| Q_A|2^1\vec P\rangle\,\langle 2^1\vec P|\vec T_A|2^3\vec P\rangle\,
\lbr \frac{1}{3}\,\vec I\cdot\vec L - I^i\, L^j\, (S^i\,S^j)^{(2)}\rbr
\nonumber \\&& 
+ \langle 2^3 \vec P|\hat Q_A|2^1\vec P\rangle\,
\langle 2^1\vec P|\vec T_A|2^3\vec P\rangle\,
\lbr -\frac{1}{6}\,\vec I\cdot \vec L
+ \frac{9}{20}\,I^i\,S^j\,(L^i\,L^j)^{(2)}
+ \frac{1}{20}\,I^i\,L^j\,(S^i\,S^j)^{(2)}\rbr \biggr]\,,
\end{eqnarray}
\end{widetext}
where the $Q_A$ and $\vec{T}_A$ operators include the anomalous magnetic moment
and the recoil additions,
\begin{align}
& Q_A =   (1+a_e)\, \frac{Z}{3}\, 4\pi \bigl[
   \delta^3(r_1)-\delta^3(r_2)\bigr]\,, \\
& \vec{T}_A = (1+2a_e)\, \frac{Z}{4}\, \left[ \frac{\vec{r}_1}{r_1^3}\times \vec{p}_1
  - \frac{\vec{r}_2}{r_2^3}\times \vec{p}_2 \right]
  + \frac{1}{4}\, \frac{\vec{r}}{r^3} \times (\vec{p}_2+\vec{p}_1)
 \nonumber \\ & \ \ \ \ \  \ \ \ \ \
  + \frac{m}{M}\,(1+a_e)\, \frac{Z}{2}
   \left(\frac{\vec{r}_1}{r_1^3} - \frac{\vec{r}_2}{r_2^3}\right) \times (\vec{p}_1+\vec{p}_2)
\,,
\end{align}
and the matrix elements are calulated between the wave functions that 
include the mass polarization correction.

The higher-order mixing contribution is obtained from Eq.~(\ref{e124}) after
subtracting the part that is already accounted for 
[namely, the $n=2$ term 
in Eq.~(\ref{singletP})].
Numerical results for the corresponding matrix elements 
are given in Table~\ref{tab:mixho}. 
The recoil correction is due to both the
the mass polarization and the recoil addition to $T_A$.
Further corrections to this mixing (e.g., those
coming from higher-order relativistic corrections) are not known and
contribute to the uncertainty of final results.

\begin{table}
\caption{
Nuclear-spin dependent fine-structure 
mixing matrix elements, with the anomalous magnetic moment and
mass polarization (upper entry) and without (lower entry), in a.u.,
all figures shown are exact,
$(A,B) \equiv \left\langle 2^3P| A | 2^1P\right\rangle \, 
\left\langle 2^1P| B| 2^3P\right\rangle
/[E(2^3P)-E(2^1P)].$
The expected numerical uncertainty is less than 1 on the last significant
digit.
\label{tab:mixho}
} 
\begin{ruledtabular}
  \begin{tabular}{l.}
$({Q_A}$ , $\vec{T}_A)$     & 156.x1127 \\
                            & 155.x9034 \\
$(\hat{Q}_A$ , $\vec{T}_A)$ &  -0.x81535 \\
                            &  -0.x81428 
  \end{tabular}
\end{ruledtabular}
\end{table}

\section{Results and discussion}

In this section we present the total theoretical predictions for the mixed fine
and hyperfine structure of the $2^3P$ states in $^3$He. The
numerical values of nuclear parameters used in the calculations are \cite{nist:web}
\begin{eqnarray}
\mu/\mu_B &=& -1.158\,740\,958(14)\,10^{-3}\,, \\
m/M &=&        1.819\,543\,0761(17)\,10^{-4}\,.
\end{eqnarray}
The conversion factors relevant for this work are
\begin{eqnarray}
\frac{m_r^3}{m\,M}(1+\kappa)\,\alpha^4 &=& -202\,887.3247\ {\rm kHz} \times h\,, \label{const}\\
\frac{m_r^3}{m\,M}(1+\kappa)\,\alpha^6 &=&   -10.8040 \ {\rm kHz}\times h\,, \\
\frac{m_r^5}{m^2\,M^2}(1+\kappa)^2\,\alpha^6 &=& 0.0063 \ {\rm kHz}\times h\,,
\end{eqnarray}
where $h$ is the Planck constant.

The fine structure splitting in the absence of the nuclear spin was calculated in
our previous investigations \cite{pachucki:09:hefs,pachucki:10:hefs}, with
numerical results reported for $^4$He. In this
work, we reevaluate all nuclear-mass-dependent corrections to the fine
structure, in order to extend our calculation to $^3$He. 
The numerical results for the $2^3P_J$ levels of $^3$He are 
\begin{align}
{\langle H_{\rm fs} \rangle}_{J=0} &\ = \frac{h}{9}\, (8\,f_{01} + 5\, f_{12})\,,\\
{\langle H_{\rm fs} \rangle}_{J=1} &\ = \frac{h}{9}\, (-f_{01} + 5\, f_{12})\,,\\
{\langle H_{\rm fs} \rangle}_{J=2}  &\ = \frac{h}{9}\, (-f_{01} - 4\, f_{12})\,, \label{06}
\end{align} 
where
\begin{align}
f_{01} =  29\,616\,676.5~(1.7)~\mbox{\rm kHz}\,, & \label{07}\\
f_{12} =   2\,292\,167.6~(1.7)~\mbox{\rm kHz}\,. & \label{08}
\end{align}

\begin{table}
\caption{
Theoretical results for individual $2^3P_J^{F}$ levels 
in $^3$He, relative to the $2^3P$ centroid energy,
in comparison with available experimental data, in kHz.
The first error of the theoretical values is the
uncertainty due to the hyperfine structure 
and the second error is the uncertainty
due to the fine structure.
\label{tab:hfslevel}
} 
\begin{ruledtabular}
  \begin{tabular}{l.l}
$2^3P_0^{F=1/2}$ &  27\,923x\,393.7\,(0.2)(2.5) & \\
                &  27\,923x\,394.7\,(2.2) & Smiciklas \cite{smiciklas:03} \\
                &  27\,923x\,398.3\,(1.9) & Cancio {\em et al.} \cite{cancio:12:3he} \\[2ex]
$2^3P_2^{F=3/2}$ &      498x\,547.3\,(1.4)(0.4) & \\
                &      498x\,543.7\,(2.1) & Smiciklas \cite{smiciklas:03} \\
                &      498x\,547.3\,(2.1) & Cancio {\em et al.} \cite{cancio:12:3he}\\[2ex]
$2^3P_1^{F=1/2}$ &     -169x\,462.2\,(0.5)(0.8) & \\
                &     -169x\,463.3\,(1.7) & Smiciklas \cite{smiciklas:03} \\
                &     -169x\,460.2\,(1.8) & Cancio {\em et al.} \cite{cancio:12:3he}\\[2ex]
$2^3P_1^{F=3/2}$ &  -4\,681x\,676.3\,(0.7)(0.2) & \\
                &  -4\,681x\,676.3\,(1.5) & Smiciklas \cite{smiciklas:03} \\
                &  -4\,681x\,672.1\,(1.6) & Cancio {\em et al.} \cite{cancio:12:3he}\\[2ex]
$2^3P_2^{F=5/2}$ &  -6\,462x\,557.8\,(0.7)(1.0) & \\
                &  -6\,462x\,555.3\,(1.5) & Smiciklas \cite{smiciklas:03} \\
                &  -6\,462x\,562.8\,(1.6) & Cancio {\em et al.} \cite{cancio:12:3he}
  \end{tabular}

\end{ruledtabular}
\end{table}


\begin{table}
\caption{
Experimental and theoretical hyperfine transitions, in kHz.
The first error in the theoretical prediction is the
uncertainty due to the hyperfine splitting and the second error is the uncertainty
due to the fine-structure splitting.
\label{tab:hfstrans}
} 
\begin{ruledtabular}
  \begin{tabular}{l.l}
$(J,F)-(J',F')$ &  \multicolumn{1}{c}{Value} & Reference \\
\hline\\
$(0,1/2)-(1,1/2)$ & 28\,092\,855x.9\,(0.5)\,(1.7)  \\
                  & 28\,092\,870x\,(60) & Morton {\em et al.} \cite{morton:06} \\
                  & 28\,092\,858x\,(3)  & Smiciklas \cite{smiciklas:03} \\
                  & 28\,092\,858x.6\,(2.1)\footnote{
Averaged value of the two differences between the measured 
optical transitions.} & Cancio {\em et al.} \cite{cancio:12:3he} \\[2ex]
$(0,1/2)-(2,3/2)$ & 27\,424\,846x.4\,(1.4)\,(2.9) \\
                  & 27\,424\,837x\,(12)   & Wu\&Drake \cite{wu:07} \\
                  & 27\,424\,851x.0\,(3.0) & Cancio {\em et al.} \cite{cancio:12:3he}\\[2ex]
$(2,3/2)-(1,1/2)$ &     668\,009x.5\,(1.4)\,(1.2) \\
                  &     668\,033x\,(9)  &  Wu\&Drake \cite{wu:07}  \\
                  &     668\,007x\,(3)  & Smiciklas \cite{smiciklas:03} \\
                  &     668\,007x.5\,(3.2) & Cancio {\em et al.} \cite{cancio:12:3he}\\[2ex] 
$(1,3/2)-(2,5/2)$ &   1\,780\,881x.5\,(1.0)\,(1.2) \\ 
                  &   1\,780\,880x\,(1) & Wu\&Drake  \cite{wu:07} \\
                  &   1\,780\,879x\,(3)    & Smiciklas \cite{smiciklas:03} \\
                  &   1\,780\,890x.7(3.5) &  Cancio {\em et al.} \cite{cancio:12:3he} \\[2ex]
$(1,1/2)-(1,3/2)$ &   4\,512\,214x.1\,(0.8)\,(0.5) \\
                  &   4\,512\,191x\,(12) & Wu\&Drake  \cite{wu:07} \\
                  &   4\,512\,213x\,(3)  & Smiciklas \cite{smiciklas:03} \\
                  &   4\,512\,211x.9\,(2.7) & Cancio {\em et al.} \cite{cancio:12:3he}\\[2ex] 
$(2,3/2)-(2,5/2)$ &   6\,961\,105x.1\,(1.5)\,(0.5) 
  \end{tabular}
\end{ruledtabular}
\end{table}

We now have all contributions to the elements of the Hamiltonian matrix 
$E^F_{J,J'}$ in Eq.~(\ref{01}). Diagonalizing the matrix, we obtain the
positions of the energy levels of the $2^3P_J^{F}$ states in $^3$He, 
relative to the centroid of the $2^3P$ level. The numerical results are 
listed in Table~\ref{tab:hfslevel} for the individual energy levels
and in Table~\ref{tab:hfstrans} for the transitions between the fine and
hyperfine levels. Our theoretical values have two
uncertainties, the first one being due to the hyperfine effects and the second
one, due to the fine-structure effects. 
The nuclear-spin dependent effects are
calculated with an accuracy of about 1~kHz (and even better in some cases). This
accuracy is limited mainly by the incomplete treatment of the 
second-order hyperfine correction. 
The second uncertainty of the theoretical energies is due to the
fine-structure effects. It comes from the 1.7~kHz error of $f_{01}$ and
$f_{12}$ in Eqs.~(\ref{07}) and (\ref{08}), which is exactly the same as
for the fine structure in $^4$He. Interestingly, the sensitivity of different
levels to the error of the fine-structure effects is very much different,
varying from 0.2~kHz for the $2^3P_{1}^{3/2}$ level to 2.5~kHz for $2^3P_{0}^{1/2}$.
As could be expected, the transitions between the levels with the same value of $J$
are less sensitive to the error of the fine-structure effects than the $J-J'$ transitions.

The present theoretical results can be compared with the experiment by
Smiciklas~\cite{smiciklas:03} and 
with our recently reported absolute frequency measurements of 
the $2^3S$-$2^3P$ transitions \cite{cancio:12:3he}.
The latter experiment was carried out by using the optical frequency comb
assisted multi-resonant precision spectroscopy \cite{consolino:11} and by measuring
simultaneously both optical and microwave hyperfine transition frequencies. 
Agreement found between the microwave frequencies and the difference of the
optical transition frequencies was used as a confirmation of the obtained
experimental results. Comparison of the two independent measurements 
(see Table~\ref{tab:hfstrans}) shows good agreement in all cases except for 
the $P_1^{3/2}$-$P_2^{5/2}$ transition.

The experimental values for individual energy levels listed in
Table~\ref{tab:hfslevel} are obtained from the transition energies reported in
the original references \cite{smiciklas:03,cancio:12:3he} by using the
definition of the centroid [Eq.~(\ref{ce1})] and the experimental hyperfine 
shift of the $2^3S$ state \cite{schluesser:69}. We observe that theoretical
and experimental results are at the same level of accuracy of about 2-3~kHz and in
very good agreement with each other.
It can be concluded that our calculation of the $m\alpha^6$ correction represented
an important advance over the
previous theory \cite{morton:06,wu:07} and
significantly improved agreement between theory and the experiment. 

Table~\ref{tab:hfs2S1} shows our
numerical results for the hyperfine splitting of the $2^3S$ state of $^3$He.
The results listed represent an update of the calculation described in 
detail in Ref.~\cite{pachucki:01:jpb}. As compared to that work,
we have (i) slightly improved the numerical accuracy and 
(ii) made the treatment of the second-order hyperfine and the nuclear-structure 
contributions to be fully consistent with that for the $2^3P$
states. The uncertainty of the theoretical prediction comes from the
second-order hyperfine correction and was estimated in the same way as for the $2^3P$ state.
Very good agreement of the $2^3S$ theoretical result with the experimental value
\cite{schluesser:69} gives us additional confidence in
our estimation of errors for the $2^3P$ state.

\begin{figure}
\includegraphics[width=8 cm]{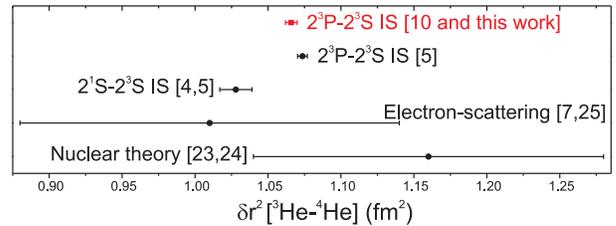}
\caption{\label{radii_fig} (Color online) 
Different determinations of
the difference of the squared nuclear charge radii of $^3$He
and $^4$He.}
\end{figure}

Having calculated the hyperfine and fine structure of the $2^3P$ levels, we
are now in a position to obtain an improved determination
of $^3$He-$^4$He nuclear charge radii difference $\delta r^2$ from the isotope shift
measurement by Shiner {\em et al.}~\cite{shiner:95}. In order to extract 
$\delta r^2$ from the measured energy difference,
$E(^3{\rm He},2^3P_0^{1/2}$-$2^3S_1^{3/2})$--$E(^4{\rm He},2^3P_2$-$2^3S_1)$,
we subtract the experimental hyperfine shift of the $2^3S$ state
\cite{schluesser:69}, the theoretical shift
of the $2^3P_0^{1/2}$ level with respect to the centroid energy (obtained in this
work), the theoretical fine shift of the $2^ 3P_2$ level with respect to the centroid 
\cite{pachucki:10:hefs}, and the theoretical isotope shift of the centroids for the
point nucleus \cite{cancio:12:3he} (see Table~\ref{tab:rms}).   
The remainder $\delta E$ comes from the finite nuclear size effect and is proportional to 
the difference of the mean square charge radii, $\delta E = C\,\delta r^2$,
where coefficient $C$ is evaluated in this work to be
$C = -1212.2(1)$ kHz/fm$^2$. The resulting value
\begin{align}
\delta r^2 \equiv r^2(^3{\rm He}) - r^2(^4{\rm He}) = 1.066(4)\; {\rm fm}^2
\end{align} 
is in reasonable agreement with the result by Cancio {\rm et al.}
\cite{cancio:12:3he}, 
$\delta r^2 = 1.074(3)$ fm$^2$,
and in significant disagreement with the result of Ref.~\cite{rooij:11}
(updated in Ref.~\cite{cancio:12:3he} by recalculation of the isotope shift), 
$\delta r^2 = 1.028(11)$ fm$^2$. Figure \ref{radii_fig} shows graphically 
the comparison of different determinations of $\delta r^2$, 
including the results from nuclear theory \cite{drake:05,sick:08} 
and from nuclear electron-scattering measurements \cite{sick:08:1,kievsky:08}.

\begin{table}
\caption{
Hyperfine splitting of the $2^3S$ state of $^3$He, in kHz.
\label{tab:hfs2S1}
} 
\begin{ruledtabular}
  \begin{tabular}{l.}
$H^{(4+)}$         & 6\,740\,451.x46 \\
$H^{(6)}$          &     -1\,313.x99 \\
$(H^{(4)}_{\rm hfs},H^{(4)}_{\rm nfs}+H^{(4)}_{\rm fs})$&        
                         2\,189.x81 \\
$(H^{(4)}_{\rm hfs},H^{(4)}_{\rm hfs})$& 
                            -60.x52 \pm 1.7 \\
$H_{\rm nucl\&ho}$ &     -1\,566.x83 \\
Total            &  6\,739\,699.x93 \pm 1.7 \\
Experiment \cite{rosner:70}      &  6\,739\,701.x177\,(16) \\
  \end{tabular}
\end{ruledtabular}
\end{table}


\begin{table*}
\caption{Determination of the difference of the mean square nuclear charge
radii $\delta r^2$
of $^3$He and $^4$He from the measurement by Shiner {\em et al.}~\cite{shiner:95}.
The remainder $\delta E$ is proportinal to $\delta r^2$,
$\delta E = C\,\delta r^2$, with $C = -1212.2(1)$ kHz/fm$^2$,
see text for details.  Units are kHz.
\label{tab:rms}
} 
\begin{ruledtabular}
  \begin{tabular}{l.l}
$E(^3{\rm He},2^3P_0^{1/2} - 2^3S_1^{3/2}) - E(^4{\rm He},2^3P_2 - 2^3S_1)$ &  810x\,599.(3.) &Experiment \cite{shiner:95}\\
$\delta E_{\rm hfs}(2^3S_1^{3/2})$&-2\,246x\,567.059(5) & Experiment \cite{schluesser:69}\\
$\delta E_{\rm fs}(2^3P_2)$ &-4\,309x\,074.2(1.7) & Theory,
    \cite{pachucki:10:hefs} and this work, Eq.~(\ref{06})\\
$-\delta E_{\rm hfs}(2^3P_0^{1/2})$ &-27\,923x\,393.7\,(1.7) & Theory, this
    work, Table~\ref{tab:hfslevel} \\
$-\delta E_{\rm iso}(2^3P - 2^3S)$ (point nucleus) &33\,667x\,143.2(3.9) & Theory \cite{cancio:12:3he} \\
$\delta E$                     &-1x\,292.8(5.2) & 
  \end{tabular}

\end{ruledtabular}
\end{table*}


\section{Summary}

In summary, we have calculated the mixed hyperfine and fine structure of the
$2^3P$ states of $^3$He. Our investigation advances the previous theory by a
complete calculation of the $m\alpha^6$ correction, which leads to an
order-of-magnitude improvement in accuracy. Theoretical predictions for most
of the transitions are accurate to better than 2~kHz. The 
$2^3P_1^{1/2}$-$2^3P_1^{3/2}$ transition is calculated up to 0.9~kHz, 
which is currently
the most precise theoretical result for the helium transitions. 
Both the hypefine and fine-structure transitions are in good
agreement with measurements by Cancio {\em et al.} \cite{cancio:12:3he}
and by Smiciklas \cite{smiciklas:03}.

Since the present theoretical accuracy for the fine-structure transitions in 
$^3$He is comparable to that for $^4$He, one can in principle use the $^3$He spectroscopy
for the determination of the fine structure constant $\alpha$, as in was done 
for $^4$He in Ref.~\cite{pachucki:10:hefs}. However, such determination 
does not bring much
advantage at present, since the experimental accuracy for $^3$He is lower than
for $^4$He \cite{borbely:09,smiciklas:10}. 
Because of this, we do not determine $\alpha$ from the $^3$He transitions 
in this work. However, if the experimental precision of the $^3$He hyperfine
structure is improved to the sub-kHz level, one will be able to use these results for
the spectropic determination of the fine structure constant.  

Another application of the hyperfine structure measurements in $^3$He might be
determination of the dipole magnetic moment of helion. The principal
problem here is that the present theory obtains the nuclear-structure contribution
from the experimental value of the 
hyperfine splitting in $^3$He$^+$. This greatly reduces the sensitivity of the final
theoretical prediction on the nuclear magnetic moment. Our calculation shows
that at present, the hyperfine structure of $^3$He allows for determination of 
the dipole magnetic moment of helion with an accuracy of about $5\times
10^{-5}$ only.


\appendix

\begin{widetext}

\section{Angular-momentum matrix elements}
\label{app:A}

In this section we calculate the matrix elements of the basic 
angular-momentum operators that are relevant for the present work.
The angular-momentum algebra is conveniently done with formulas
from Ref.~\cite{varshalovich}. The results are
\begin{align}
\langle FIJLSM_F|\vec{I}\cdot\vec{S}|FIJ'L'S'M_F\rangle &\  = 
(-1)^{I+J+F}\, \SixJ{I}{J}{F}{J'}{I}{1} \, \sqrt{I(I+1)(2I+1)}\,
  \nonumber \\ & \times
 \delta_{L,L'}\,\delta_{S,S'}\,(-1)^{L+S+J+1}\,
 \sqrt{S(S+1)(2S+1)(2J+1)(2J'+1)}\, 
 \SixJ{S}{L}{J'}{J}{1}{S}\,,
\end{align}
\begin{align}
\langle FIJLSM_F|\vec{I}\cdot\vec{L}|FIJ'L'S'M_F\rangle &\  = 
(-1)^{I+J+F}\, \SixJ{I}{J}{F}{J'}{I}{1} \, \sqrt{I(I+1)(2I+1)}\,
  \nonumber \\ & \times
 \delta_{L,L'}\,\delta_{S,S'}\,(-1)^{L+S+J'+1}\,
 \sqrt{L(L+1)(2L+1)(2J+1)(2J'+1)}\, 
 \SixJ{L}{S}{J}{J'}{1}{L}\,,
\end{align}
\begin{align}
\langle I^i\,S^j\,(L^i\,L^j)^{(2)}\rangle = \frac14\left[
 J(J+1)+J'(J'+1)-2 L(L+1)- 2S(S+1)\right] \,\langle
  \vec{I}\cdot\vec{L}\rangle
- \frac13\, L(L+1)\,\langle  \vec{I}\cdot\vec{S}\rangle\,,
\end{align}
\begin{align}
\langle I^i\,L^j\,(S^i\,S^j)^{(2)}\rangle = \frac14\left[
 J(J+1)+J'(J'+1)-2 L(L+1)- 2S(S+1)\right] \,\langle
  \vec{I}\cdot\vec{S}\rangle
- \frac13\, S(S+1)\,\langle  \vec{I}\cdot\vec{L}\rangle\,,
\end{align}
\begin{align}
\langle (S^i\,S^j)^{(2)} (L^i\,L^j)^{(2)}\rangle = 
 \langle \vec{L}\cdot\vec{S}\rangle^2
 + \frac12\,\langle \vec{L}\cdot\vec{S}\rangle
 -\frac{L\,(L+1)\,S\,(S+1)}{3}\,.
\end{align}

\section{Derivation of $H_{\rm hfs}^{(6)}$}
\label{app:B}

We start with the Breit Hamiltonian $H_{BP}$ of the atomic system in the
external magnetic field,
\begin{eqnarray}
H_{BP} &=& \sum_a H_a + \sum_{a,b; a>b} H_{ab}\,\\
H_a &=& \frac{\vec\pi_a^2}{2\,m}-\frac{Z\,\alpha}{r_a}-\frac{e}{2\,m}\,
\vec\sigma_a\cdot\vec B_a-\frac{\vec\pi_a^4}{8\,m^3}+
\frac{\pi\,Z\,\alpha}{2\,m^2}\,\delta^3(r_a)
+\frac{Z\,\alpha}{4\,m^2}\,\vec\sigma_a\cdot\frac{\vec r_a}{r_a^3}\times\vec\pi_a
\nonumber \\ &&
+\frac{e}{8\,m^3}\,\bigl(\vec\sigma_a\cdot\vec B_a\,\vec \pi_a^2 +
\vec \pi_a^2\,\vec\sigma_a\cdot\vec B_a\bigr)\,,
\\
H_{ab} &=& \frac{\alpha}{r_{ab}} +\frac{\pi\,\alpha}{m^2}\,\delta^3(r_{ab})
-\frac{\alpha}{2\,m^2}\,\pi_a^i\,\biggl(\frac{\delta^{ij}}{r_{ab}}
+\frac{r_{ab}^i\,r_{ab}^j}{r_{ab}^2}\biggr)\,\pi_b^j
\nonumber \\ &&
+\frac{\alpha}{4\,m^2r_{ab}^3}\bigl[
\vec\sigma_a\cdot\vec r_{ab}\times(2\,\vec\pi_b-\vec\pi_a)
-\vec\sigma_b\cdot\vec r_{ab}\times(2\,\vec\pi_a-\vec\pi_b)\bigr]
\nonumber \\ &&
+\frac{\alpha}{4\,m^2}\,\frac{\sigma_a^i\,\sigma_b^j}{r_{ab}^3}\,
\biggl(\delta^{ij}-3\,\frac{r_{ab}^i\,r_{ab}^j}{r_{ab}^2}\biggr)\,,
\end{eqnarray}
where $\vec\pi = \vec p-e\,\vec A$. 
Magnetic fields $\vec A$ and $\vec B$ induced by the nuclear magnetic moment are
\begin{eqnarray}
e\,\vec A(\vec r) &=& \frac{e}{4\,\pi}\,\vec\mu\times\frac{\vec r}{r^3} =
-Z\,\alpha\,\frac{(1+\kappa)}{M}\,\vec I\times\frac{\vec r}{r^3}\,,\label{B4}\\
e\,\vec B^i(\vec r) &=& \bigl(\vec \nabla\times\vec A\bigr)^i =
-Z\,\alpha\,\frac{(1+\kappa)}{M}\,\frac{8\,\pi}{3}\,\delta^3(r)\,I^i
+Z\,\alpha\,\frac{(1+\kappa)}{M}\,\frac{1}{r^3}\,\biggl(\delta^{ij}-
3\,\frac{r^i\,r^j}{r^2}\biggr)\,I^j\,.\label{B5}
\end{eqnarray}

The leading-order interaction 
between the nuclear spin $\vec I$ and the electron spin $\vec \sigma_a$
is obtained from the nonrelativistic terms
$\nicefrac1{2\,m}\,\vec\pi_a^2$ and $\nicefrac{e}{2\,m}\,\vec\sigma_a\cdot\vec B_a$,
yielding
\begin{align}
H^{(4)}_{\rm hfs} = -\sum_a \biggl[ \frac{e}{m}\,\vec p_a\cdot\vec A(\vec r_a)
-\frac{e}{2\,m}\,\vec\sigma_a\cdot\vec B(\vec r_a)\biggr]\,,
\end{align}
in agreement with Eq.~(\ref{09}).
The relativistic correction to the hyperfine interaction is similarly obtained
from the relativistic terms in the Breit Hamiltonian $H_{BP}$,
\begin{eqnarray}
H^{(6)}_{\rm hfs} &=& \sum_a
\frac{Z\,\alpha}{4\,m^2}\,\vec\sigma_a\cdot\frac{\vec
  r_a}{r_a^3}\times\bigl[-e\,\vec A(\vec r_a)\bigr]+
\frac{e}{8\,m^3}\,\bigl(\vec\sigma_a\cdot\vec B_a\,\vec p_a^{\,2} +
\vec p_a^{\,2}\,\vec\sigma_a\cdot\vec B_a\bigr)\nonumber \\ &&
+\sum_{a,b; a\neq b} \frac{\alpha}{4\,m^2r_{ab}^3}\,
\vec\sigma_a\cdot\vec r_{ab}\times
\bigl[-2\,e\,\vec A(\vec r_b)+e\,\vec A(\vec r_a)\bigr]\label{B7}\\&&
+\frac{e}{4\,m^3}\,\sum_a \bigl[
\vec p_a^{\;2}\,\vec p_a\cdot\vec A(\vec r_a)+
\vec p_a\cdot\vec A(\vec r_a)\,\vec p_a^{\;2}\bigr]
-\sum_{a\neq b}\frac{\alpha}{2\,m^2}\,p_a^i\,\biggl(\frac{\delta^{ij}}{r_{ab}}
+\frac{r_{ab}^i\,r_{ab}^j}{r_{ab}^3}\biggr)\,
\bigl[-\,e\,A^j(\vec r_b)\bigr]\,.
\end{eqnarray}
Using $\vec A$ and $\vec B$ from Eqs. (\ref{B4}) and (\ref{B5}), we derive Eq.~(\ref{eH6}).

\section{Second order matrix elements}
\label{app:C}

The second-order $m\alpha^6$ correction $\delta E_{\rm sec}$ [see Eq.~(\ref{e123})]
is split into
several parts in accordance with the symmetry of the intermediate states,
\begin{align}
& \delta E_{\rm sec} = 
\frac{m_r^3}{m\,M}\,(1+\kappa)\,\alpha^6\,\biggl[
\delta E_{\rm sing}(^3P) 
+ \delta E_{\rm reg}(^3P) 
+ 
\delta E(^1P) + \delta E(^3D) + \delta E(^1D) + \delta E(^3F)\biggr]\,.\label{26}
\end{align}
The angular momentum algebra for different intermediate states is simple but rather tedious
and is performed with help of the symbolic algebra computer program. Below we
list the resulting formulas
expressed in a form convenient for the numerical evaluation. 

The singular part with the $^3P$ intermediate states is given by
(after removing all divergencies) 
\begin{align}
\delta E_{\rm sing}(^3P) =
 \biggl\langle Q'\,\frac{1}{(E-H)'}\,T'\biggr\rangle\, 2\, \lbr \vec
 I\cdot\vec S \rbr\,,
\end{align}
where
$E\equiv E(2^3P)$,
\begin{eqnarray}
Q' &\equiv& 
Q+
\frac{2}{3}\,\sum_a\,\biggl\{\frac{Z}{r_a}\,,\,E-H\biggr\}
= -\frac{2\,Z}{3}\,\biggl(
\frac{\vec r_1}{r_1^3}\cdot\vec \nabla_1
+\frac{\vec r_2}{r_2^3}\cdot\vec \nabla_2\biggr) \,, \\
  T' & \equiv & T
-\frac{1}{4}\,\sum_a\,\biggl\{\frac{Z}{r_a}\,,\,E-H\biggr\}
=  -\frac{1}{2}\,\left(E+\frac{Z}{r_1}+\frac{Z}{r_2}-\frac1{r}\right)^2
  -p_1^i\,\frac{1}{2\,r}\,\biggl(\delta^{ij}+\frac{r^i\,r^j}{r^2}\biggr)\,p_2^j
  \nonumber \\ && 
\ \ \ \ \ \ \ \ \ \ \ \ \ \ \ \ \ \ \ \ \ \ \ \ \ \ \ \ \ \ \ \ \ \ \ 
+\frac{1}{4}\,\vec \nabla_1^2 \, \vec \nabla_2^2
  -\frac{Z}{4}\,\frac{\vec r_1}{r_1^3}\cdot\vec \nabla_1
  -\frac{Z}{4}\,\frac{\vec r_2}{r_2^3}\cdot\vec \nabla_2 
   +\frac{1}{2}\,\frac{\vec r}{r^3}\cdot(\vec\nabla_1-\vec\nabla_2)\,,\label{40}
\end{eqnarray}
and it is assumed that the operators $T'$ and $Q'$ act on the function on the right hand side
that satisfies the Schr\"odinger equation with the energy $E$.

The regular part with the $^3P$ intermediate states is
\begin{eqnarray}
\delta E_{\rm reg}(^3P) &=&\sum_{n>2}\,\frac{1}{E(2^3P)-E(n^3P)}\biggl[
\langle 2^3\vec P|Q|n^3 \vec P\rangle\,\langle n^3 \vec P|\vec T|2^3 \vec P\rangle\,
\lbr\frac{2}{3}\,\vec I\cdot\vec L+I^i\,L^j\,(S^i\,S^j)^{(2)}\rbr
\nonumber \\&&
+\langle 2^3\vec P|Q|n^3 \vec P\rangle\,
\langle n^3 \vec P|\hat T|2^3 \vec P\rangle\,\lbr-\frac{3}{5}\,I^i\,S^j\,(L^i\,L^j)^{(2)}\rbr
+\langle 2^3\vec P|\vec Q|n^3\vec P\rangle\,
\langle n^3\vec P|T|2^3\vec P\rangle\,\lbr \vec I\cdot\vec L \rbr
\nonumber \\ &&
+\langle 2^3 \vec P|\vec Q|n^3 \vec P\rangle\,\langle n^3\vec P|\vec T|2^3 \vec P\rangle\,
\lbr\frac{1}{3}\,\vec I\cdot\vec S+\frac{1}{2}\,I^i\,S^j\,(L^i\,L^j)^{(2)}\rbr
\nonumber \\&&
+\langle 2^3 \vec P|\vec Q|n^3 \vec P\rangle\,\langle n^3\vec P|\hat T|2^3 \vec
P\rangle\,\lbr-\frac{3}{10}\,I^i\,L^j\,(S^i\,S^j)^{(2)}\rbr 
\nonumber \\&&
+\langle 2^3 \vec P|\hat Q|n^3 \vec P\rangle\,
\langle n^3 \vec P|T|2^3 \vec P\rangle\,\lbr-\frac{6}{5}\,\,I^i\,S^j\,(L^i\,L^j)^{(2)}\rbr
\nonumber \\&&
+\langle 2^3 \vec P|\hat Q|n^3 \vec P\rangle\,\langle n^3 \vec P|\vec T|2^3 \vec P\rangle\,
\lbr-\frac{1}{3}\,\vec I\cdot\vec L+
\frac{9}{20}\,I^i\,S^j\,(L^i\,L^j)^{(2)}-\frac{1}{20}\,I^i\,L^j\,(S^i\,S^j)^{(2)}\rbr
\nonumber \\&&
+\langle 2^3\vec P|\hat Q|n^3 \vec P\rangle\,\langle n^3\vec P|\hat T|2^3 \vec P\rangle\,
\lbr\frac{1}{5}\,\vec I\cdot\vec S -
\frac{21}{100}\,I^i\,S^j\,(L^i\,L^j)^{(2)}-\frac{27}{100}\,I^i\,L^j\,(S^i\,S^j)^{(2)}\rbr
\biggr]\,.
\end{eqnarray}

The other parts are 
\begin{align} \label{singletP}
\delta E(^1P) = &\ 
\sum_{n}\,\frac{1}{E(2^3P)-E(n^1P)}\biggl[
\langle 2^3\vec P| Q_A|n^1\vec P\rangle\,\langle n^1\vec P|\vec T_A|2^3\vec P\rangle\,
\lbr\frac{1}{3}\,\vec I\cdot\vec L - I^i\, L^j\, (S^i\,S^j)^{(2)}\rbr
\nonumber \\& 
+ \langle 2^3 \vec P|\hat Q_A|n^1\vec P\rangle\,
\langle n^1\vec P|\vec T_A|2^3\vec P\rangle\,
\lbr-\frac{1}{6}\,\vec I\cdot \vec L
+ \frac{9}{20}\,I^i\,S^j\,(L^i\,L^j)^{(2)}
+ \frac{1}{20}\,I^i\,L^j\,(S^i\,S^j)^{(2)}\rbr\biggr]\,,
\\
\delta E(^3D) &\ =
\sum_{n}\,\frac{1}{E(2^3P)-E(n^3D)}\biggl[
\langle 2^3\vec P|\vec Q|n^3\hat D\rangle\,
\langle n^3\hat D|\vec T|2^3\vec P\rangle\,\lbr
\frac{2}{3}\,\vec I\cdot \vec S - \frac{1}{5}\,I^i\,S^j\,(L^i\,L^j)^{(2)}\rbr
\nonumber \\&
+ \langle 2^3\vec P|\hat Q|n^3\hat D\rangle\,
\langle n^3\hat D|\hat T|2^3\vec P\rangle\,
\lbr\frac{2}{9}\,\vec I\cdot\vec S
+ \frac{7}{30}\,I^i\,S^j\,(L^i\,L^j)^{(2)}
- \frac{1}{10}\,I^i\,L^j\,(S^i\,S^j)^{(2)}\rbr
\nonumber \\&
+ \langle 2^3\vec P|\hat Q|n^3\hat D\rangle\,
\langle n^3\hat D|\vec T|2^3\vec P\rangle\,
\lbr -\frac{2}{3}\,\vec I\cdot\vec L -
\frac{3}{10}\,I^i\,S^j\,(L^i\,L^j)^{(2)}  -
\frac{1}{10}\, I^i\,L^j\,(S^i\,S^j)^{(2)}\rbr
\nonumber \\&
+ \langle 2^3\vec P|\vec Q|n^3\hat D\rangle\,
\langle n^3\hat D|\hat T|2^3\vec P\rangle\,
\,\lbr -\frac{3}{5}I^i\,L^j\,(S^i\,S^j)^{(2)}\rbr\biggr]\,,\\
\delta E(^1D) &\ =
\sum_{n}\,\frac{1}{E(2^3P)-E(n^1D)}\,
\langle 2^3\vec P|\hat Q_A|n^1\hat D\rangle\,
\langle n^3\hat D|\vec T_A|2^1\vec P\rangle
\lbr-\frac{1}{3}\,\vec I\cdot\vec L
- \frac{3}{10}\,I^i\,S^j\,(L^i\,L^j)^{(2)}
+ \frac{1}{10}\,I^i\,L^j\,(S^i\,S^j)^{(2)}\rbr\,,\\
\delta E(^3F) &\ =
\sum_{n}\,\frac{1}{E(2^3P)-E(n^3F)}\,
\langle 2^3\vec P|\hat Q|n^3\tilde F\rangle\,
\langle n^3\tilde F|\hat T|2^3\vec P\rangle
\lbr\frac{1}{3}\,\vec I\cdot\vec S
- \frac{1}{10}\,I^i\,S^j\,(L^i\,L^j)^{(2)}
+ \frac{3}{10}\,I^i\,L^j\,(S^i\,S^j)^{(2)}\rbr\,.
\end{align}

The matrix elements with the $D$ and $F$-state wave functions
are defined by
\begin{eqnarray} \label{17}
\langle\vec\phi|\vec Q|\hat\psi\rangle &\equiv& -i\,\langle\phi_i|Q_j|\psi_{ij}\rangle
                                   = i\,\langle\psi_{ij}|Q_j|\phi_i\rangle               \equiv \langle\hat\psi|\vec Q|\vec\phi\rangle\,,\\
\langle\vec\phi|\hat Q|\hat\psi\rangle &\equiv& \epsilon^{ijk}\langle\phi_i|Q_{jl}|\psi_{kl}\rangle
                                   = \epsilon^{ijk}\langle\psi_{kl}|Q_{jl}|\phi_i\rangle \equiv \langle\hat\psi|\hat Q|\vec\phi\rangle\,,\\
\langle\vec\phi|\hat Q|\tilde\psi\rangle &\equiv& \langle\phi_i|Q_{jk}|\psi_{ijk}\rangle
                                   = \langle\psi_{ijk}|Q_{ij}|\phi_k\rangle               \equiv \langle\tilde\psi|\hat Q|\hat\phi\rangle
\,,\label{22}
\end{eqnarray}
where 
$\hat{\psi}$ denotes the odd $D$ wave function,
\begin{align}
\psi^{ij} = \left( \epsilon^{ikl} r_1^k r_2^l r_1^j +
    \epsilon^{jkl} r_1^k r_2^l r_1^i \right) \, f(r_1,r_2,r) \pm (r_1 \leftrightarrow r_2)\,,
\end{align}
and $\tilde{\psi}$ denotes the odd $F$ wave function,
\begin{align}
\psi^{ijk} = &\ \left[ r_1^ir_1^jr_1^k - \frac{r_1^2}{5} \left( \delta_{ij}r_1^k
  +\delta_{ik}r_1^i + \delta_{jk}r_1^i\right)\right]\,f(r_1,r_2,r)
   + \frac13\biggl[r_1^i r_1^j r_2^k +r_1^i r_2^j r_1^k + r_2^i r_1^j r_1^k
   \nonumber \\ &
- \frac{\delta_{ij}}{5} (r_1^2 r_2^k +2 r_1^lr_2^l  r_1^k)
- \frac{\delta_{ik}}{5} (r_1^2 r_2^j +2 r_1^lr_2^l  r_1^j)
- \frac{\delta_{jk}}{5} (r_1^2 r_2^i +2 r_1^lr_2^l  r_1^i)
\biggr] \,g(r_1,r_2,r)
 \pm (r_1 \leftrightarrow r_2)\,.
\end{align}
All wave functions are real and normalized by $\langle \psi^i|\psi^i \rangle = 
\langle \psi^{ij}|\psi^{ij} \rangle = \langle \psi^{ijk}|\psi^{ijk} \rangle = 1$.

The second-order corrections listed above are finite. However, some matrix elements
are too singular for the direct numerical evaluation and need to be
transformed to a more regular form. The regularization method is described in
Sec.~\ref{sec:ma6}. So, the operator $T$ is transformed by Eq.~(\ref{e125}),
thus yielding
\begin{align}
&\sum_{n>2}\,\frac{1}{E(2^3P)-E(n^3P)}
\langle 2^3\vec P|\vec Q|n^3\vec P\rangle\,
\langle n^3\vec P|T|2^3\vec P\rangle  
   \nonumber\\ &
= \sum_{n>2}\,\frac{1}{E(2^3P)-E(n^3P)}
\langle 2^3\vec P|\vec Q|n^3\vec P\rangle\,
\langle n^3\vec P|T'|2^3\vec P\rangle 
+\langle 2^3\vec P|\frac{1}{4}\,\sum_a\frac{Z}{r_a}\,\vec Q\,|2^3\vec P\rangle
-\langle 2^3\vec P|\frac{1}{4}\,\sum_a\frac{Z}{r_a}|2^3\vec P\rangle\,
\langle 2^3\vec P|\vec Q\,|2^3\vec P\rangle\,.
\end{align}
We observe that, after the transformation, the singular part of the operator
$T$ is absorbed in the first-order matrix element, whereas the second-order
correction contains only the (more regular) operator $T'$. 

Similarly, the operator $Q^{ij}$ is transformed by
\begin{eqnarray}
Q'^{ij} &\equiv& Q^{ij}-\bigl\{\delta Q^{ij}\,,\,E-H\bigr\}
= -\frac{Z}{6}\,\sum_a\biggl(
-\delta^{ij}\,\frac{r_a^k}{r_a^3}-3\,\delta^{ik}\,\frac{r_a^j}{r_a^3}
-3\,\delta^{jk}\,\frac{r_a^i}{r_a^3} + 9\,\frac{r_a^i\,r_a^j\,r_a^k}{r_a^5}\biggr)\,\nabla_a^k\,,
\end{eqnarray}
where
$E\equiv E(2^3P)$,
\begin{align}
\delta Q^{ij} = \frac{1}{6}\,\sum_a\,
\frac{Z}{r_a}\,\biggl(\delta^{ij}-3\frac{r_a^i\,r_a^j}{r_a^2}\biggr)\,
\end{align}
and it is assumed that the function on the right hand side of $\hat Q'$ 
satisfies the Schr\"odinger align with energy $E$.
The second-order matrix elements with $Q^{ij}$ then are transformed by
\begin{align}
&\sum_{n>2}\,\frac{1}{E(2^3P)-E(n^3P)}
\langle 2^3\vec P|\hat Q|n^3\vec P\rangle\,
\langle n^3\vec P|\vec T|2^3\vec P\rangle  
   \nonumber\\ &
= \sum_{n>2}\,\frac{1}{E(2^3P)-E(n^3P)}
\langle 2^3\vec P|\hat Q'|n^3\vec P\rangle\,
\langle n^3\vec P|\vec T|2^3\vec P\rangle 
+\left( -i \epsilon^{klm}\right) \langle 2^3 P_i|\delta Q_{ik}\,T_l\,|2^3 P_m\rangle
-\langle 2^3\vec P|\delta \hat Q|2^3\vec P\rangle\,
\langle 2^3\vec P|\vec T\,|2^3\vec P\rangle\,,
\end{align}
and
\begin{align}
&\sum_{n>2}\,\frac{1}{E(2^3P)-E(n^3P)}
\langle 2^3\vec P| \hat Q|n^3\vec P\rangle\,
\langle n^3\vec P|T|2^3\vec P\rangle 
= \sum_{n>2}\,\frac{1}{E(2^3P)-E(n^3P)}
\langle 2^3\vec P| \hat Q'|n^3\vec P\rangle\,
\langle n^3\vec P|T'|2^3\vec P\rangle \nonumber \\
&+\langle 2^3\vec P|\frac{1}{4}\,\sum_a\frac{Z}{r_a}\,\hat Q\,|2^3\vec P\rangle
-\langle 2^3\vec P|\frac{1}{4}\,\sum_a\frac{Z}{r_a}|2^3\vec P\rangle\,\langle 2^3\vec P|\hat Q|2^3\vec P\rangle
+\langle 2^3\vec P|\delta \hat Q\,T'|2^3\vec P\rangle
-\langle 2^3\vec P|\delta \hat Q|2^3\vec P\rangle\,\langle 2^3\vec P|T'|2^3\vec P\rangle\,.
\end{align}

The operator $Q_A$ is transformed as
\begin{eqnarray} \label{Qap}
Q'_A &\equiv& Q_A+\frac{2}{3}\,
\biggl\{\frac{Z}{r_1}-\frac{Z}{r_2}\,,\,E-H\biggr\}
= \frac{2\,Z}{3}\,
\biggl(-\frac{\vec r_1}{r_1^3}\cdot\vec\nabla_1 + \frac{\vec r_2}{r_2^3}\cdot\vec\nabla_2\biggr)\,,
\end{eqnarray}
where it is assumed that the function on the right hand side of $Q'_A$ 
satisfies the Schr\"odinger equation with the energy $E$. The regularized form of the second-order
matrix element then is
\begin{eqnarray}
&&\sum_{n}\,\displaystyle{\frac{1}{E(2^3P)-E(n^1P)}}\,
\langle 2^3\vec P| Q_A|n^1\vec P\rangle\,\langle n^1\vec P|\vec T_A|2^3\vec P\rangle
=
\nonumber \\ &&
\sum_{n}\,\displaystyle{\frac{1}{E(2^3P)-E(n^1P)}}\,
\langle 2^3\vec P| Q'_A|n^1\vec P\rangle\,\langle n^1\vec P|\vec T_A|2^3\vec P\rangle
-\frac{2}{3}\,\langle 2^3\vec P|\left(\frac{Z}{r_1}-\frac{Z}{r_2}\right)
\vec T_A|2^3\vec P\rangle\,.
\end{eqnarray}

\end{widetext}

\end{document}